\documentclass[sigconf, screen]{acmart}

\author{Celeste Barnaby}
\affiliation{%
  \institution{Facebook Inc.}
  \country{U.S.A.}
}
\email{celestebarnaby@fb.com}

\author{Koushik Sen}
\affiliation{%
 \institution{UC Berkeley}
 \country{U.S.A.}
}
\email{ksen@berkeley.edu}

\author{Tianyi Zhang}
\affiliation{%
 \institution{Harvard University}
 \country{U.S.A.}
}
\email{tianyi@seas.harvard.edu}

\author{Elena Glassman}
\affiliation{%
  \institution{Harvard University}
  \country{U.S.A.}
}
\email{glassman@seas.harvard.edu}

\author{Satish Chandra}
\affiliation{%
  \institution{Facebook Inc.}
  \country{U.S.A.}
}
\email{schandra@acm.org}

\AtBeginDocument{%
  \providecommand\BibTeX{{%
    \normalfont B\kern-0.5em{\scshape i\kern-0.25em b}\kern-0.8em\TeX}}}

\usepackage[]{algorithmic, algorithm}
\usepackage{graphicx}
\usepackage{fancyvrb}
\usepackage{endnotes}
\usepackage{hyperref}
\usepackage{subfig}
\usepackage{multirow}
\usepackage{enumitem}
\usepackage{soul}
\usepackage{microtype}
\graphicspath{ {./} }
\hypersetup{
    colorlinks=true,
    linkcolor=blue,
}

\DeclareFixedFont{\ttbi}{T1}{txtt}{bx}{n}{10} 
\DeclareFixedFont{\ttmi}{T1}{txtt}{m}{n}{10}  
\DeclareFixedFont{\ttb}{T1}{txtt}{bx}{n}{8} 
\DeclareFixedFont{\ttm}{T1}{txtt}{m}{n}{8}  
\DeclareFixedFont{\ttbsmall}{T1}{txtt}{bx}{n}{7.5} 
\DeclareFixedFont{\ttmsmall}{T1}{txtt}{m}{n}{7.5}  

\usepackage{color}
\definecolor{deepblue}{rgb}{0,0,0.5}
\definecolor{deepred}{rgb}{0.6,0,0}
\definecolor{deepgreen}{rgb}{0,0.5,0}
\definecolor{deeppurple}{rgb}{0.6,0,0.5}

\usepackage{listings}

\newcommand{\todo}[1]{\textcolor{red}{\textsf{TODO}: #1}}

\newcommand\pythonstyle{\lstset{
language=Python,
basicstyle=\ttm,
otherkeywords={self},             
keywordstyle=\ttb\color{deepblue},
commentstyle=\ttm\color{deeppurple},
emph={Select, Subset, OrderedSubset, Sequence, Chain},          
emphstyle=\ttb\color{deepred},    
stringstyle=\color{deepgreen},
frame=tb,                         
showstringspaces=false,            %
numbers=left,
numberstyle=\ttmsmall,
numbersep=2pt,
xleftmargin=1.25em,
stepnumber=1,
escapechar=@
}}

\newcommand\pythonstylesmall{\lstset{
language=Python,
basicstyle=\ttmsmall,
otherkeywords={self},             
keywordstyle=\ttbsmall\color{deepblue},
commentstyle=\ttmsmall\color{deeppurple},
emph={Select, Subset, OrderedSubset, Sequence, Chain},          
emphstyle=\ttbsmall\color{deepred},    
stringstyle=\color{deepgreen},
frame=tb,                         
showstringspaces=false,            %
numbers=left,
numberstyle=\ttmsmall,
numbersep=2pt,
xleftmargin=1.5em,
stepnumber=1,
escapechar=@
}}

\lstnewenvironment{python}[1][]
{
\pythonstyle
\lstset{#1}
}
{}
\lstnewenvironment{pythonsmall}[1][]
{
\pythonstylesmall
\lstset{#1}
}
{}

\newcommand\pythoninline[1]{{\pythonstyle\lstinline!#1!}}

\newlength\myindent
\newcommand\bindent{%
  \begingroup
  \setlength{\itemindent}{\myindent}
  \addtolength{\algorithmicindent}{\myindent}
}
\newcommand\eindent{\endgroup}

\newlist{questions}{enumerate}{2}
\setlist[questions,1]{label=\textbf{RQ\arabic*.},ref=RQ\arabic*}
\setlist[questions,2]{label=(\alph*),ref=\thequestionsi(\alph*)}

\setcopyright{rightsretained}
\acmPrice{}
\acmDOI{10.1145/3368089.3417052}
\acmYear{2020}
\copyrightyear{2020}
\acmSubmissionID{fse20ind-p36-p}
\acmISBN{978-1-4503-7043-1/20/11}
\acmConference[ESEC/FSE '20]{Proceedings of the 28th ACM Joint European Software Engineering Conference and Symposium on the Foundations of Software Engineering}{November 8--13, 2020}{Virtual Event, USA}
\acmBooktitle{Proceedings of the 28th ACM Joint European Software Engineering Conference and Symposium on the Foundations of Software Engineering (ESEC/FSE '20), November 8--13, 2020, Virtual Event, USA}

\begin{document}
\sloppy

\title{Exempla Gratis (E.G.): Code Examples for Free}


\begin{abstract}

Modern software engineering often involves using many existing APIs, both open source and -- in industrial coding environments -- proprietary. Programmers reference documentation and code search tools to remind themselves of proper common usage patterns of APIs. However, high-quality API usage examples are computationally expensive to curate 
and maintain, and API usage examples retrieved from company-wide code search can be tedious to review. We present a tool, \aroma{}, that mines codebases and shows the common, idiomatic usage examples for API methods. \aroma{} was integrated into Facebook's internal code search tool for the Hack language and evaluated on open-source GitHub projects written in Python. \aroma{} was also compared against code search results and hand-written examples from a popular programming website called ProgramCreek. 
Compared with these two baselines, examples generated by \aroma{} are more succinct and representative with less extraneous statements. In addition, a survey with Facebook developers shows that \aroma{} examples are preferred in 97\% of cases.

\end{abstract}



\newcommand{\aroma}{\textsc{EG}}

\maketitle





\section{Introduction}
\label{seC:intro}

Application programming interfaces (APIs) are becoming a pervasive component of modern software engineering. A core challenge for software engineers in industry is to use existing APIs in idiomatic ways within their organization. In order to do this, developers often search for API documentation and \textit{usage examples}
~\cite{duala2012asking, robillard2009makes, buse2012synthesizing}. However, this can be especially challenging in companies where many APIs are proprietary. Because those proprietary APIs are only documented within the company by its engineers, there is no externally crowdsourced documentation or examples posted on sites like StackOverflow. 



Usage examples are a key component of API documentation~\cite{maalej2013patterns}. Examples can refresh a programmer's memory~\cite{brandt2009two}, concretize the more abstract components of documentation~\cite{robillard2009makes}, and support code improvement and adaptation~\cite{zhang2019analyzing, luan2019aroma}. However, there is some risk that programmers will generalize from or adapt an example incorrectly, e.g., leaving in irrelevant components or leaving out critical ones. One useful trait of an example is succinctness~\cite{nasehi2012makes}: having minimal details specific to a particular usage situation and few superficial distractions, leaving just what is common across most or all proper usages of the API. Alternatively, multiple examples showing a variety of usages~\cite{variationtheory} may help programmers infer what may be common and uncommon usage patterns and parameter values. Writing usage examples that have these helpful properties is labor-intensive, especially when there exist multiple proper, consistently used usage patterns within an organization.

Regardless of whether available documentation includes one or more usage examples, many programmers instead use company- or project-wide code search to find API usage snippets. 
However, code search engines results ranking is difficult and often defaults to showing the most recently edited files first. The task of sifting through myriad code search results in an attempt to glean a common usage pattern can be tedious, time-consuming, and unproductive~\cite{starke2009working}.
 If the developer does decide to use code from a code search result, they have no assurance that this code represents a common usage, rather than an atypical, niche way of using a method. In fact, prior work has shown that individual code examples may even suffer from API usage violations~\cite{zhang2018code}, insecure coding practices~\cite{fischer2017stack}, and unchecked obsolete usage~\cite{zhou2016api}. Therefore, without thoroughly inspecting and comparing many examples, developers may leave out critical safety checks or desirable usage scenarios.

Several approaches have been previously proposed to address this challenge of presenting programmers with good API usage examples, whether found through search or automatically generated. A number of approaches cluster and rank similar examples to reduce the cognitive load of reading through individual examples~\cite{kim2010towards, buse2012synthesizing, Katirtzis2018clustering}. However, these clustering techniques rely on pre-defined similarity metrics and do not help users understand why some examples are clustered together, e.g., the commonalities and variations among those examples. 
Buse et al.~ presented a synthesis technique to generate a single example from multiple similar examples~\cite{buse2012synthesizing}, but the synthetic example only demonstrates a common skeleton, without showing possible variations.
In contrast, Examplore is capable of visualizing an entire distribution over API usage features in a large number of API usage examples~\cite{glassman2018visualizing}, but the analysis requires a pre-defined API skeleton and concrete usage patterns are best revealed through additional interaction with the visualization. 

In this paper, we present \aroma{}, a tool that mines codebases and shows multiple common, idiomatic usage examples for API methods. \aroma{} assumes access to a large repository of Python programs from many projects.  Given a query API method, \aroma{} first searches all methods in the repository containing at least one usage of the API method.  It then computes the parse tree of each such method and finds the maximal subtree that a) contains the query API, and b) is part of a meaningful proportion of methods. \aroma{} then serializes the subtree to create a common idiomatic usage pattern of the API method. \aroma{} repeats this process multiple times to find $n$ diverse idiomatic usage patterns. 

Once \aroma{} has generated several common usage patterns, it displays the patterns to  users in an easy-to-use interface. For each usage pattern, \aroma{} displays a concise and representative code snippet to serve as an example of that usage pattern. In each example, \aroma{} emphasizes the code parts that are part of the common usage pattern in bold texts, while graying out the uncommon parts -- referred to in this paper as "filler". Further, \aroma{} displays how many times each usage pattern appears in the repository. This interface allows users to efficiently understand the common usage of an API method, and relieves the cognitive load of manually looking through code results in an attempt to discern a common pattern. In additional, \aroma{} relieves the burden of manually curating examples for API methods, and automates the task of keeping API examples up-do-date and relevant as a codebase changes. 



\aroma{} has several properties that are particularly advantageous for its scalability and generality. First, \aroma{} is language agnostic: to generate \aroma{} examples for a new programming language, one need only implement a new parser. Second, 
\aroma{} does not require mining coding patterns ahead of time, and can retrieve new and idiomatic usage patterns on-the-fly. 
Third, \aroma{} is fast enough to use in real time, and can generate examples from a large corpus containing millions of methods within a couple of seconds on a multi-core server machine. On average, \aroma{} takes 1.0
seconds to generate examples for a query method on a 24-core CPU.


We have implemented \aroma{} in C++ for Hack and Python. We have also integrated \aroma{} into Facebook's internal code search website, where it is used daily by developers. We report our experimental evaluation of \aroma{} for Python. We have used \aroma{} to index 1,900,911 Python methods obtained from open source GitHub projects. We performed our experiments for Python because it is a language that is widely used at Facebook, as well as in open source projects. We evaluated \aroma{} against code search results and examples from ProgramCreek, a website providing code examples of Python methods. We found that developers preferred \aroma{} examples to code search results in over 99\% of cases, and that a majority of developers found the main features of the \aroma{} interface useful. We also found that \aroma{} examples were shorter, more relevant, and more representative than code search results or ProgramCreek examples. 

The rest of this paper is organized as follows. Section~\ref{sec:motivation} motivates the design of \aroma{} with insights and lessons learned from deploying another code search and recommendation tool in Facebook. Section~\ref{sec:usage} describes a usage scenario of learning APIs with \aroma{}. Section~\ref{sec:approach} describes the pattern mining and example generation algorithms in \aroma{}. Section~\ref{sec:evaluation} describes the evaluation of \aroma{}, including a survey with Facebook developers, a quantitative analysis of examples generated by \aroma{} and two other tools, and a summary of \aroma{}'s usage metrics after its deployment in Facebook. Section~\ref{sec:discussion} discusses the challenges we encountered when evaluating \aroma{}. Section~\ref{sec:related} discusses related work and Section~\ref{sec:conclusion} concludes this paper.

\begin{table*}\scriptsize
\caption{\aroma{} code examples for a variety of Python methods.}
\label{tab:intro-examples2}

\setlength{\tabcolsep}{0.01\textwidth}
\begin{tabular}{@{}| c | p{0.38\textwidth} | p{0.32\textwidth} |@{}}
\toprule
Query & Examples & Notes \\
\midrule
\emph{Case A}: \texttt{json.dump}
&
\vspace{0.4em}
\emph{This usage pattern is found in 29 out of 336 samples. \endnote{This code snippet is adapted from \url{https://github.com/openai/gym/blob/master/gym/wrappers/monitoring/video_recorder.py\#L229}. Accessed in March 2020.} }
\begin{Verbatim}[commandchars=\\\{\}]
\textbf{with open(}self.output_path\textbf{, 'w') as f:}
 \textbf{   \hl{json.dump}(}data\textbf{, f)}
\end{Verbatim}
\rule{0.38\textwidth}{.5pt}
\emph{This usage pattern is found in 17 out of 336 samples.
\endnote{This code snippet is adapted from \url{https://github.com/scrapinghub/splash/blob/master/scripts/rst2inspections.py\#L77}. Accessed in March 2020.} }
\begin{Verbatim}[commandchars=\\\{\}]
\textbf{with open(}out_filename\textbf{, "w") as f:}
        \textbf{\hl{json.dump}(}info\textbf{, f, indent=}2\textbf{)}
\end{Verbatim}
\rule{0.38\textwidth}{.5pt}
\emph{This usage pattern is found in 17 out of 336 samples.
\endnote{This code snippet is adapted from \url{https://github.com/supernnova/SuperNNova/blob/master/supernnova/utils/experiment_settings.py\#L161}. Accessed in March 2020.}
}
\begin{Verbatim}[commandchars=\\\{\}]
\textbf{with open(}Path(self.rnn_dir) / "cli_args.json"\textbf{,} "w"\textbf{) as} f:
    \textbf{\hl{json.dump}(}self.cli_args\textbf{,} f\textbf{, indent=}4\textbf{, sort_keys=True)}
\end{Verbatim}
&
The first example shows that the following is idiomatic:
\begin{itemize}
    \item Opening a file before calling \texttt{json.dump}
    \item Passing \texttt{'w'} as the second argument to \texttt{open}
    \item Passing \texttt{f} as the second argument to \texttt{json.dump}
\end{itemize}
The second and third examples show that it is also idiomatic to pass an integer to the optional parameter \texttt{indent}, and to pass \texttt{True} to the optional parameter \texttt{sort\_keys}
\\

\midrule
\emph{Case B}: \\ \texttt{os.makedirs}
&
\emph{This usage pattern is found in 103 out of 1699 samples.
\endnote{This code snippet is adapted from \url{https://github.com//huggingface/transformers/tree/master/examples/run_multiple_choice.py}. Accessed in March 2020.}
}
\begin{Verbatim}[commandchars=\\\{\}]
output_dir \textbf{= os.path.join(}args.output_dir,
                    "checkpoint-{}".format(global_step))
\textbf{if not os.path.exists(}output_dir\textbf{):}
    \textbf{\hl{os.makedirs}(}output_dir\textbf{)}
\end{Verbatim}
\rule{0.38\textwidth}{.5pt}
\emph{This usage pattern is found in 110 out of 1699 samples.
\endnote{This code snippet is adapted from \url{https://github.com/zalandoresearch/fashion-mnist/blob/master/configs.py\#L49}. Accessed in March 2020.}
}
\begin{Verbatim}[commandchars=\\\{\}]
base_dir \textbf{= os.path.dirname(}fname\textbf{)}
\textbf{if not os.path.exists(}base_dir\textbf{):}
    \textbf{\hl{os.makedirs}(}base_dir\textbf{)}
\end{Verbatim}
\rule{0.38\textwidth}{.5pt}
\emph{This usage pattern is found in 116 out of 1699 samples.
\endnote{This code snippet is adapted from \url{https://github.com/toddheitmann/PetroPy/blob/master/petropy/download.py\#L195}. Accessed in March 2020.}
}
\begin{Verbatim}[commandchars=\\\{\}]
year_dir = \textbf{os.path.}join\textbf{(}save_dir,
                    url.split('/')[-1].split('.')[0]\textbf{)}
\textbf{if not os.path.isdir(}year_dir\textbf{):}
    \textbf{\hl{os.makedirs}(}year_dir\textbf{)}
\end{Verbatim}
&
    The first example shows that the following is idiomatic:
    \begin{itemize}
        \item Calling \texttt{os.path.join} and \texttt{os.path.exists} before calling \texttt{os.makedirs}.
        \item Calling \texttt{os.makedirs} on the condition that the directory you are making does not already exist.
    \end{itemize}
    The second example shows an alternate idiom where \texttt{os.path.dirname} is called instead of \texttt{os.path.join}, while the third example calls \texttt{os.path.isdir} instead of \texttt{os.path.exists}. 
 \\
\midrule
\emph{Case C}: \texttt{range}
&
\vspace{0.4em}
\emph{This usage pattern is found in 213 out of 2000 samples.
\endnote{This code snippet is adapted from \url{https://github.com/TarrySingh/Artificial-Intelligence-Deep-Learning-Machine-Learning-Tutorials/blob/master/deep-learning/1-pixel-attack/networks/capsnet.py\#L41}}
}
\begin{Verbatim}[commandchars=\\\{\}]
\textbf{for i in \hl{range}(}3\textbf{):}
    img[:,:,i] \textbf{=} (img[:,:,i] - mean[i]) / std[i]
\end{Verbatim}
\rule{0.38\textwidth}{.5pt}
\emph{This usage pattern is found in 150 out of 2000 samples.
\endnote{This code snippet is adapted from \url{https://github.com/scikit-learn-contrib/category_encoders/blob/master/category_encoders/sum_coding.py\#L238}. Accessed in March 2020.}
}
\begin{Verbatim}[commandchars=\\\{\}]
columns\textbf{=[}str(col) + '_\%d' \% (i, ) 
    \textbf{for i in \hl{range}(}len(sum_contrast_matrix.column_suffixes)\textbf{)]}
\end{Verbatim}
\rule{0.38\textwidth}{.5pt}
\emph{This usage pattern is found in 123 out of 2000 samples.
\endnote{This code snippet is adapted from \url{https://github.com/waditu/tushare/blob/master/tushare/util/common.py\#L40}. Accessed in March 2020.}
}
\begin{Verbatim}[commandchars=\\\{\}]
\textbf{for} j in \textbf{\hl{range}(}start\textbf{,} i\textbf{):}
\end{Verbatim}
&
    The first example shows that the following is idiomatic:
    \begin{itemize}
        \item Calling \texttt{range} in the condition of a for loop.
        \item Naming the for loop variable \texttt{i}.
    \end{itemize}
    The second example shows a common idiom for list comprehension using range, while the third example shows that two variables may be passed to \texttt{range}.
    \\
\midrule
\emph{Case D}: \\ \texttt{csv.writer}
&
\emph{This usage pattern is found in 11 out of 160 samples.
\endnote{This code snippet is adapted from \url{https://github.com/bboczeng/Nyxar/blob/master/api/coinmarketcap.py\#L94}. Accessed in March 2020.}
}
\begin{Verbatim}[commandchars=\\\{\}]
\textbf{with open(}filename\textbf{, 'a+', newline='') as} file\textbf{:}
    \textbf{writer = \hl{csv.writer}(}file\textbf{)}
    \textbf{writer.writerow(}fieldnames\textbf{)}
\end{Verbatim}
\rule{0.38\textwidth}{.5pt}
\emph{This usage pattern is found in 11 out of 160 samples.
\endnote{This code snippet is adapted from \url{https://github.com/vgpena/next-weekend/blob/master/scraper.py\#L82}}
}
\begin{Verbatim}[commandchars=\\\{\}]
\textbf{writer = \hl{csv.writer}(}csvfile, delimiter=',', 
                    quotechar='|', quoting=csv.QUOTE_MINIMAL\textbf{)}
\textbf{writer.writerow([}hike_name, url, trailhead_name, ...\textbf{])}
\end{Verbatim}
\rule{0.38\textwidth}{.5pt}
\emph{This usage pattern is found in 11 out of 160 samples.
\endnote{This code snippet is adapted from \url{https://github.com/baychimo/loto/blob/master/tests/test_loto.py\#L134}. Accessed in March 2020.}
}
\begin{Verbatim}[commandchars=\\\{\}]
\textbf{with} open(ntf.name, "w") \textbf{as} f:
    ntf_writer \textbf{= \hl{csv.writer}(}f\textbf{, delimiter=}","\textbf{)}
\end{Verbatim}
&
The first example shows that the following is idiomatic:
\begin{itemize}
    \item Opening a file before calling \texttt{csv.writer}
    \item Passing \texttt{'a+'} as the second argument to \texttt{open}, and passing \texttt{''} as the argument to the optional parameter \texttt{newline}
    \item Calling \texttt{writer.writerow} after \texttt{csv.writer}
\end{itemize}
The second example shows an alternate idiom where a list of items is passed to \texttt{writer.writerow}, while the third example shows an idiom where an argument is provided for the optional parameter \texttt{delimiter}
\\
\midrule
\emph{Case E}: \\ \texttt{requests.post}
&
\emph{This usage pattern is found in 67 out of 1019 samples.
\endnote{This code snippet is adapted from \url{https://github.com/home-assistant/core/blob/master/homeassistant/components/ohmconnect/sensor.py\#L70}. Accessed in March 2020.}
}
\begin{Verbatim}[commandchars=\\\{\}]
\textbf{response = \hl{requests.get}(}url\texttt{,} timeout\textbf{=}10\textbf{)}
\end{Verbatim}
\rule{0.38\textwidth}{.5pt}
\emph{This usage pattern is found in 48 out of 1019 samples.
\endnote{This code snippet is adapted from \url{https://github.com/zvtvz/zvt/blob/master/zvt/recorders/exchange/china_index_list_spider.py\#L85}. Accessed in March 2020.}
}
\begin{Verbatim}[commandchars=\\\{\}]
\textbf{try:}
    response \textbf{= \hl{requests.get}(}url\textbf{)}
\textbf{except} requests.HTTPError \textbf{as} error:
\end{Verbatim}
\rule{0.38\textwidth}{.5pt}
\emph{This usage pattern is found in 48 out of 1019 samples.
\endnote{This code snippet is adapted from \url{https://github.com/testerSunshine/12306/blob/master/verify/pretreatment.py\#L26}. Accessed in March 2020.}
}
\begin{Verbatim}[commandchars=\\\{\}]
\textbf{url =} 'https://kyfw.12306.cn/otn/passcodeNew/...'
\textbf{r = \hl{requests.get}(}url\textbf{)}
\end{Verbatim}
&
The first example shows that the following is idiomatic:
\begin{itemize}
    \item Naming the variable assigned to \texttt{requests.get} "\texttt{response}"
    \item Passing two arguments to \texttt{requests.get}
\end{itemize}
The second example shows an alternate idiom where the call to \texttt{requests.get} is wrapped in a try-catch block, while the third example shows that it is common to initialize a string variable names \texttt{url} before calling \texttt{requests.get}.
\\

\bottomrule

\end{tabular}
\end{table*}

\section{Motivations from Facebook}
\label{sec:motivation}
Aroma is a code-to-code search and recommendation tool. It has been integrated into Facebook's IDE and internal code search website in December 2018~\cite{luan2019aroma}. Given a code snippet as input and a large code corpus, Aroma returns a set of idiomatic extensions to the input code clustered together from similar code snippets in the corpus. Aroma produces code recommendations for Hack, Python, Java, and JavaScript. 
Here, we summarize how the lessons we learned from Aroma informed the design of \aroma{}.

We expected that developers would query Aroma with multi-line code snippets, to get recommendations for how they should modify or improve their code. 
However, we found that in practice, most Aroma queries were for single API methods. Furthermore, most of these queried APIs were Facebook-specific APIs for which there was little existing documentation and no hand-written examples. We concluded, then, that developers at Facebook were using Aroma to obtain API usage examples. 

\begin{figure}[h]
\includegraphics[scale=.5]{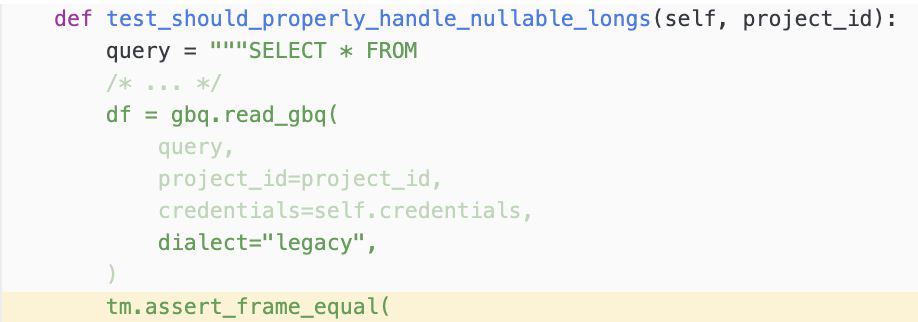}
\caption{Aroma recommendation for assert\_frame\_equal\protect\endnote{This code snippet is adapted from \url{https://github.com/pydata/pandas-gbq/blob/master/tests/system/test_gbq.py\#L130}. Accessed in March 2020.}}
\label{fig:aroma_rec}
\end{figure}

Since Aroma was not designed for generating examples of API usage, recommendations created from querying a single API method had several shortcomings. First, we found that across many different methods, APIs, and libraries, Aroma recommendations consistently cut out the arguments passed into a function call. For example, in Figure~~\ref{fig:aroma_rec}, the example generated by Aroma does not include any arguments to the \texttt{assert\_frame\_equal} method in \texttt{pandas.testing}. This is because Aroma is designed to prune out code that is different among multiple snippets in a cluster, while retaining code that is commonly shared among them. Since different calls to this method tend to contain different arguments, the arguments are pruned out in the recommendation. Second, examples generated by Aroma include many extraneous statements. In Figure~\ref{fig:aroma_rec}, this example contains several lines that are not strictly relevant to the  \texttt{assert\_frame\_equal} call, such as the function header, and the initialization of the \texttt{query} variable. 

For API learning, these shortcomings are detrimental. When learning the common usage of an API method, it is helpful to see its common arguments, and usually unhelpful to see a lot of extraneous context. Prior work has shown that conciseness is an important feature of code examples, and that the median length of hand-written examples is five lines~\cite{nasehi2012makes}. Aroma's ability to perform fuzzy searches also goes under-utilized when the query is a single method.



\begin{figure}[h]
\includegraphics[scale=.6]{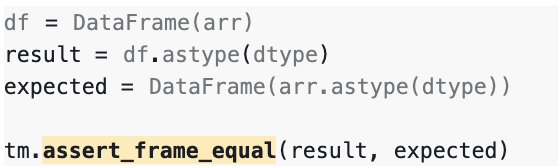}
\caption{\aroma{}'s example for \texttt{assert\_frame\_equal}
\protect\endnote{This code snippet is adapted from \url{https://github.com/pandas-dev/pandas/blob/master/pandas/tests/frame/test_dtypes.py}. Accessed in March 2020.}}
\label{fig:nametbd_example}
\end{figure}

For these reasons, we decided that, while Aroma is still a powerful code recommendation engine with other potential uses, it is not suitable as a generator of API usage examples. 
Thus, we created \aroma{} to allow developers to see succinct, idiomatic usage examples for an API method. Figure~\ref{fig:nametbd_example} shows the top example generated by \aroma{} for learning \texttt{assert\_frame\_equal}. This example includes the arguments of the queried method and removes those extraneous statements. The additional code serves only to further illuminate the use of \texttt{assert\_frame\_equals}, as it shows how to initialize its arguments. Further, \aroma{} shows code elements that are common in black text, and code elements that are unique to a single snippet in gray text. This allows users to understand what is common and what is atypical, while still seeing a complete, readable example. 

\section{Usage Scenario}
\label{sec:usage}
This section describes a usage scenario of learning Python APIs with \aroma{}. While we find \aroma{} to be most useful for proprietary libraries with few hand-written examples, we cannot release such proprietary code in this paper for confidentiality reasons. Thus, for the purposes of this scenario, we assume that hand-written examples for the libraries mentioned are not widely available. 

\begin{figure}[ht!]
\includegraphics[scale=.4]{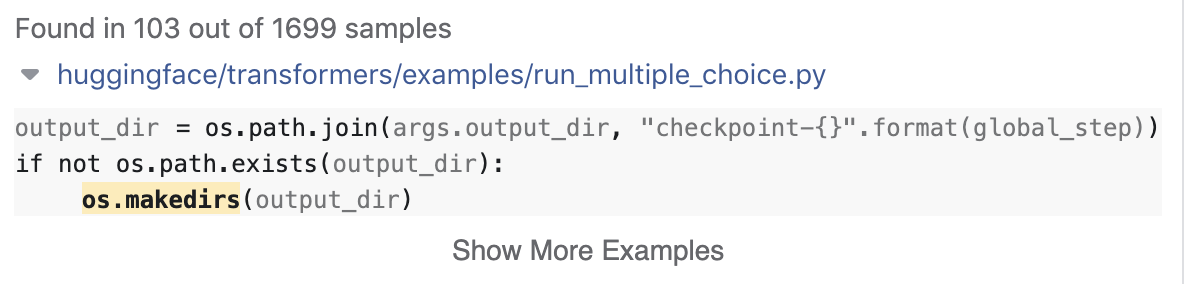}
\caption{\aroma{}'s interface showing an example for \texttt{os.makedirs}. When code search results initially load, the top \aroma{} example is presented as the first result.\protect\endnotemark[4]}
\label{fig:makedirs2}
\end{figure}

Suppose Harry is a novice Python developer. He needs to write code that creates a directory and then writes some text to a file in that directory. He is aware that there is a \texttt{makedirs} function in the \texttt{os} package, but he is not sure how to use it. He searches for \texttt{os.makedirs} in \aroma{}. 
Figure~\ref{fig:makedirs2} shows the top example generated by \aroma{}. This example shows that among 1699 snippets that call \texttt{os.makedirs}, 103 followed the same API usage pattern. The bolded code in this example shows the idiomatic usage of \texttt{os.path.exists}. Harry finds that it is common to check whether the directory exists before creating it. Further, he finds that it is idiomatic to call \texttt{os.path.join} together with \texttt{os.makedirs} to safely construct a file path across platforms. 
A link to the file containing the code snippet used in this example is displayed above the code snippet, which Harry can use if he wants to see additional context. 

\begin{figure}[h]
\includegraphics[scale=.4]{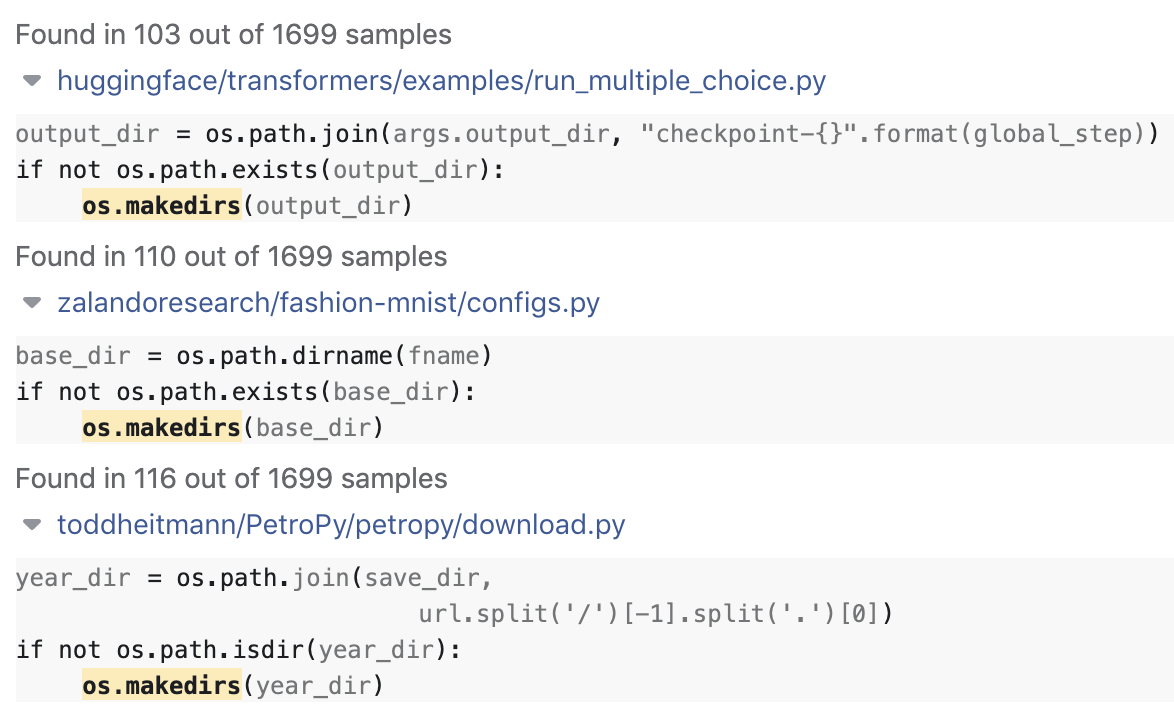}
\caption{The "Show More Examples" button displays the top three common usage examples.\protect\endnotemark[4]\protect\endnotemark[5]\protect\endnotemark[6]}
\label{fig:makedirs1}
\end{figure}

Harry clicks "Show More Examples" to view additional usage examples of \texttt{os.makedirs}, as shown in Figure \ref{fig:makedirs1}. He sees that the third example calls \texttt{os.path.isdir} in the 
\texttt{if} statement instead of \texttt{os.path.exists}. The text above this code example indicates that this is a common usage pattern appearing in 116 out of 1699 snippets, giving Harry confidence that this is another standard check before calling \texttt{os.makedirs}.
Harry copies this code from the \aroma{} example and replaces \texttt{year\_dir} with the name of his directory. Since these examples generated by \aroma{} have already summarized distinct API usage in hundreds of examples in the codebase, Harry feels he does not look at any additional code search results.  

\begin{figure}[h]
\includegraphics[scale=.4]{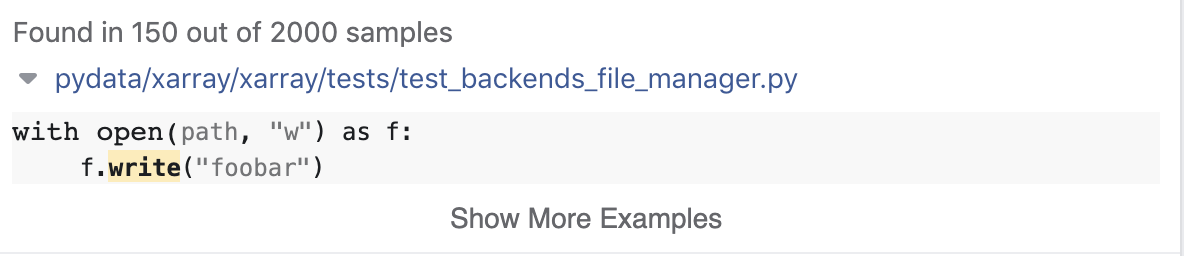}
\caption{\aroma{}'s example for \texttt{write}.\protect\endnote{This code is adapted from \url{https://github.com/pydata/xarray/blob/master/xarray/tests/test_backends_file_manager.py\#L197}. Accessed in March 2020.}}
\label{fig:write1}
\end{figure}

Harry now needs his code to write text to a file, so he queries \texttt{write} in \aroma{}, without including a package name. Figure \ref{fig:write1} shows the top example generated by \aroma{}. He sees that this common usage pattern is found in 150 methods out of 2000 snippets, indicating that it is idiomatic to open a file before calling \texttt{write}. He also sees that, in this code snippet, the second argument to \texttt{open} is \texttt{"w"}. 
Further search shows that \texttt{"w"} means write-only. This is exactly what Harry needs. By stitching this example with the previous example, Harry successfully writes the desired code. 

\section{Example generation algorithm}
\label{sec:approach}

In this section, we describe several notations and definitions to compute the simplified parse tree of a program.  The terminologies and notations are similar to that in Aroma~\cite{luan2019aroma}.  We reintroduce the definitions to keep the paper self-contained.

\subsection{Formal Definitions}

\begin{definition}[Keyword tokens] This is the set of all tokens in a language whose values are fixed as part of the language. Keyword tokens include keywords such as \texttt{while} and \texttt{if}, and symbols such as \texttt{\{}, \texttt{\}}, \texttt{.}, \texttt{+}, \texttt{*}. The set of all keyword tokens is finite for a language.
\end{definition}

\begin{definition}[Non-keyword tokens] This is the set of all tokens that are not keyword tokens.  Non-keyword tokens include variable names, method names, field names, and literals.  \end{definition}

\noindent Examples of non-keyword tokens are \texttt{i}, \texttt{length}, $0$, $1$, etc.  The set of non-keyword tokens is non-finite for most languages.

\begin{definition}[Simplified Parse Tree]  A simplified parse tree is a data structure to represent a program.  It is recursively defined as a non-empty list whose elements could be any of the following:
    \begin{itemize}
        \item a non-keyword token,
        \item a keyword token, or
        \item a simplified parse tree.
    \end{itemize}
    Moreover, a simplified parse tree cannot be a list containing a single simplified parse tree.  
\end{definition}

We picked this particular representation of programs instead of a conventional abstract syntax tree representation because the representation only consists of program tokens, and does not use any special language-specific rule names such as \texttt{IfStatement}, \texttt{block} etc.  As such, the representation can be used uniformly across various programming languages.  Moreover, one could perform an in-order traversal of a simplified parse tree and print the token names to obtain the original program.  We use this feature of a simplified parse tree to show the common usage examples.

\begin{definition}[Label of a Simplified Parse Tree]  The label of a simplified parse tree is obtained by concatenating all the elements of the list representing the tree as follows:
    \begin{itemize}
        \item If an element is a keyword token, the value of the token is used for concatenation.
        \item If an element is a non-keyword token or a simplified parse tree, the special symbol \texttt{\#} is used for concatenation.
    \end{itemize}
\end{definition}

\noindent For example, the label of the simplified parse tree \texttt{["x", ">", ["y", ".", "f"]]} is \texttt{"\#>\#"}.

\begin{figure}[!ht]
\begin{Verbatim}[commandchars=\\\{\},frame=single]
\textbf{with open(}self.output_path\textbf{,} \textbf{'w') as f:}
    \textbf{    json.dump(}data\textbf{, f)}
\end{Verbatim}
\includegraphics[scale=.35]{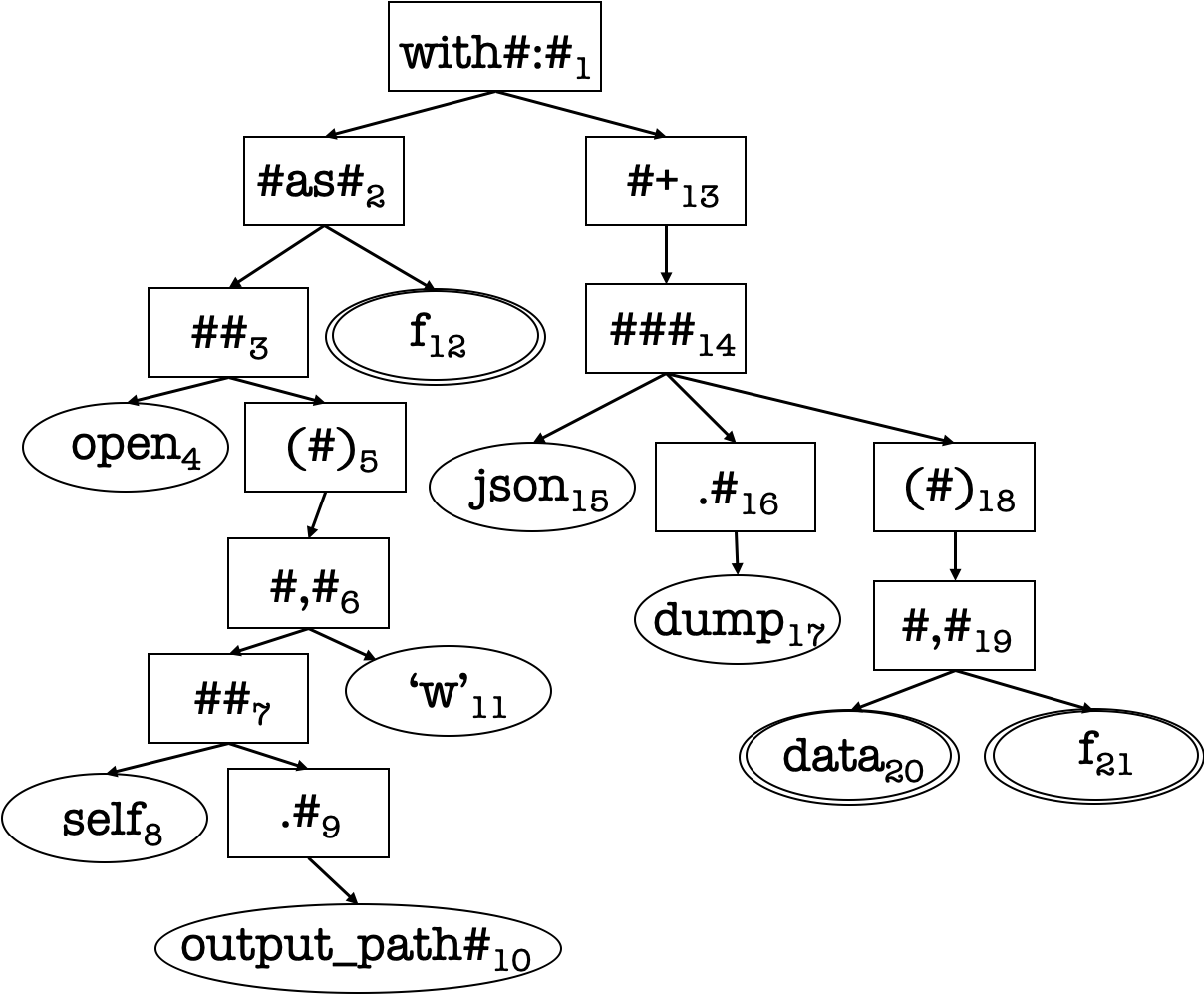}
\caption{A simplified parse tree of a code snippet. Variable nodes are highlighted in double circles.}
\label{fig:example_tree}
\end{figure}

Figure~\ref{fig:example_tree} shows a code snippet and its simplified parse tree. In the figure, each internal node represents a simplified parse tree, and is labeled using the tree's label as defined above.  Since keyword tokens in a simplified parse tree become part of the label of the tree, we do not create leaf nodes for keyword tokens in the tree diagram---we only add leaf nodes for non-keyword tokens.  We show the label of each node in the tree, and add a unique index to each label as subscript to distinguish between nodes with similar labels.

To obtain the simplified parse tree of a code snippet, \aroma{} relies on a language-specific parser. For example, \aroma{} utilizes the \texttt{lib2to3} Python parser to produce the simplified parse tree for a Python program. Once the simplified parse tree of a code snippet has been created, the rest of \aroma{}'s algorithm is language-agnostic.

We will represent a simplified parse tree $t$ using the tuple $(N, L, E)$, where
\begin{itemize}
\item $N$ is the set of nodes of the tree,
\item $L$ is a function that maps a node to the label of the subtree rooted at the node,
\item $E$ is a children function. If $n_2$ is the $i^{\rm th}$ direct child of the node $n_1$, then $E(n_1, i) = n_2$. If the $i^{\rm th}$ child of a node $n$ does not exist, then $E(n, i) = \bot$.
\end{itemize}
For example, \texttt{with\#:\#$_1$} and \texttt{self$_8$} $\in N$ are sample nodes in the tree shown in Figure~\ref{fig:example_tree}. $L($ \texttt{with\#:\#}$_1) =$ \texttt{with\#:\#}. $E($\texttt{\#as\#}$_2, 2) = $\texttt{f}$_{12}$.

\begin{figure}[!ht]
\includegraphics[scale=.35]{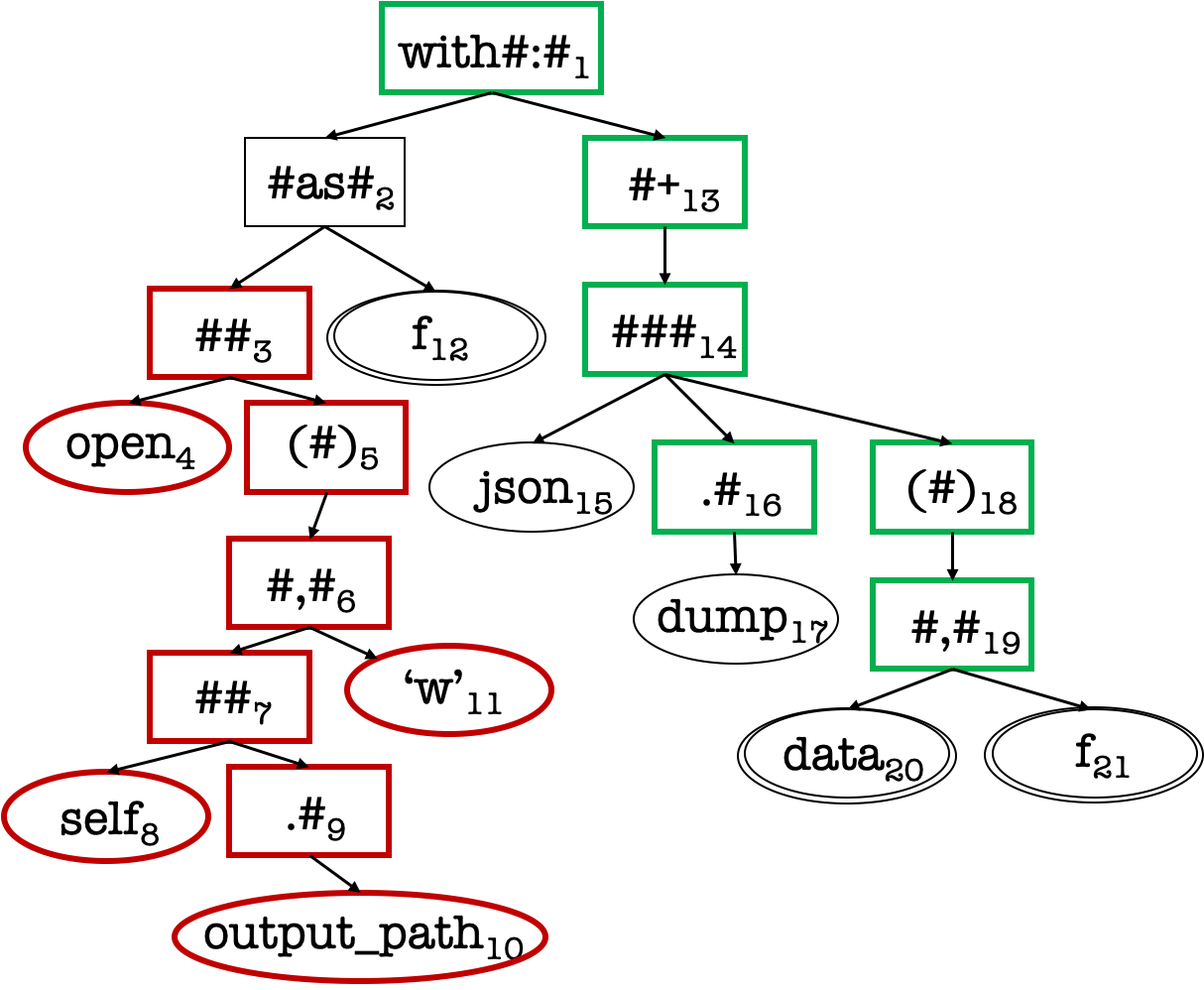}
\caption{A simplified parse tree with a subtree highlighted in red and a context highlighted in green.}
\label{fig:context_and_subtree}
\end{figure}

A \textbf{subtree} of a tree $t$ is a tree rooted at some node in $t$ and contains all the descendants of the node in $t$.  Formally, $t' = (N', L, E')$ is a subtree of $t= (N, L, E)$ if the following conditions hold:
\begin{itemize}
\item $N' \subseteq N$,
\item for all $n_1 \in N'$ if there exists $n_2 \in N$ and an $i \in \mathbb{N}$ such that $E(n_1, i) = n_2$, then $n_2 \in N'$ and $E'(n_1, i) = n_2$,
\item for all $n \in N'$, if there exists $i \in \mathbb{N}$ such that $E(n, i) = \bot$, then $E'(n, i) = \bot$.
\end{itemize}
For example, the subtree rooted at \texttt{\#\#}$_{3}$ in Figure~\ref{fig:context_and_subtree} is highlighted in red.

A \textbf{context} tree of a tree $t$ is the tree with some of its subtrees removed.  Formally, if $t' = (N', L, E')$ is a context tree of $t = (N, L, E)$, then the following conditions hold:
\begin{itemize}
\item $N' \subseteq N$,
\item for all $n_1 \in N'$ if there exists $n_2 \in N$ and a $i \in \mathbb{N}$ such that $E(n_2, i) = n_1$, then $n_2 \in N'$ and $E'(n_2, i) = n_1$.
\end{itemize}
In Figure~\ref{fig:context_and_subtree}, we highlight a context of the tree in green.

A \textbf{context subtree} can be obtained from a tree first by picking a subtree of the tree and then picking a context tree of the subtree.  We also use the term \textbf{pattern} to refer to a context subtree.

\subsection{\aroma{} Algorithm}

We assume that we are given a set of trees $T$ and a query tree $q$. Note that a query -- for example, \texttt{json.dump} -- is parsed as a code construct as well. \aroma{} works in two steps to create a common usage pattern.  First, it finds a pattern $t$ such that $q$ is a subtree of $t$ and $t$ is a pattern in each tree in a subset $T'$ of $T$.  The pattern denotes a partial code snippet that is common and contains the query snippet.  Second, \aroma{} finds a completion of the pattern by picking a subtree from a suitable tree in $T'$.  The subtree must contain the pattern as a context.  The subtree denotes a common usage code snippet of $q$.

\paragraph{Phase 1.} \aroma{} starts with the pattern $q$ and grows it iteratively by adding nodes to the pattern as shown in Figure~\ref{fig:phase1}.  Let us assume that after some iteration the current pattern is $t_c = (N, L, E)$ and it is present exactly in $T_c \subseteq T$ trees. Then a suitable neighboring node $n_1$ is added to the pattern to obtain a new bigger pattern as described in Figure~\ref{fig:extend} .  The tuple $(l, i, n_1, n_2, b)$ denotes that a node $n_1$ is added to the tree $t_c$ where $l$ is the label of $n_1$, $n_2$ is the node in $t_c$ connected to $n_1$, and $b$ is a Boolean which if true means $E(n_2,i) = n_1$, and $E(n_1,i) = n_2$ if $b$ is false. The \textbf{support} of a tree added to the pattern is the number of trees in $T_c$ that contain the new pattern (see Figure~\ref{fig:support}).  
In an iteration, \aroma{} adds a node to $t_c$ such that the new pattern has the highest support.  At the end of an iteration, \aroma{} updates $t_c$ with the new pattern and the set of all the trees in $T_c$ containing the new pattern becomes the new $T_c$.  

\aroma{} continues the iterations until the number of nodes in $t_c$ exceeds a configurable threshold $\gamma$ (usually set to 100) or the cardinality of $T_c$ divided by $T$ goes below a configurable threshold $\alpha$ (usually set to .05). Threshold $\gamma$ ensures that the generated example is not too long, while threshold $\alpha$ ensures that the generate example is a common snippet.

\begin{figure}[!tbp]
\begin{Verbatim}[commandchars=\\\{\},frame=single]
if data:
    \textbf{with open(}self.output_path\textbf{,} \textbf{'w') as f:}
        \textbf{    json.dump(}data\textbf{, f)}
\end{Verbatim}
\includegraphics[scale=.35]{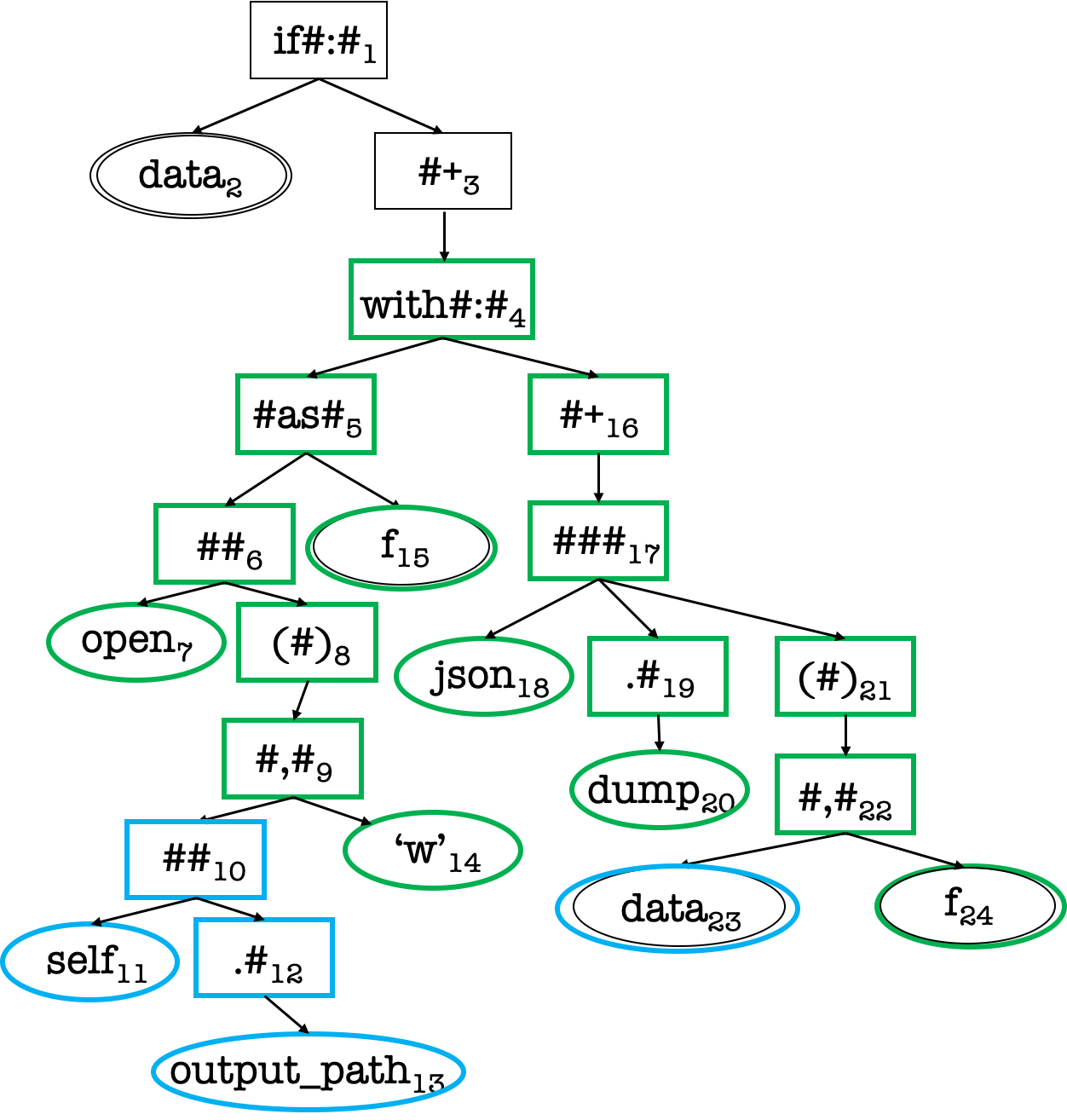}
\begin{Verbatim}[commandchars=\\\{\},frame=single]
print("Writing to \%s." \% json_path)
\textbf{with open(}json_path\textbf{,} \textbf{'w') as f:}
\textbf{   json.dump(}scan_json_results\textbf{, f)}
\end{Verbatim}
\vspace*{0.5em}
\includegraphics[scale=.3]{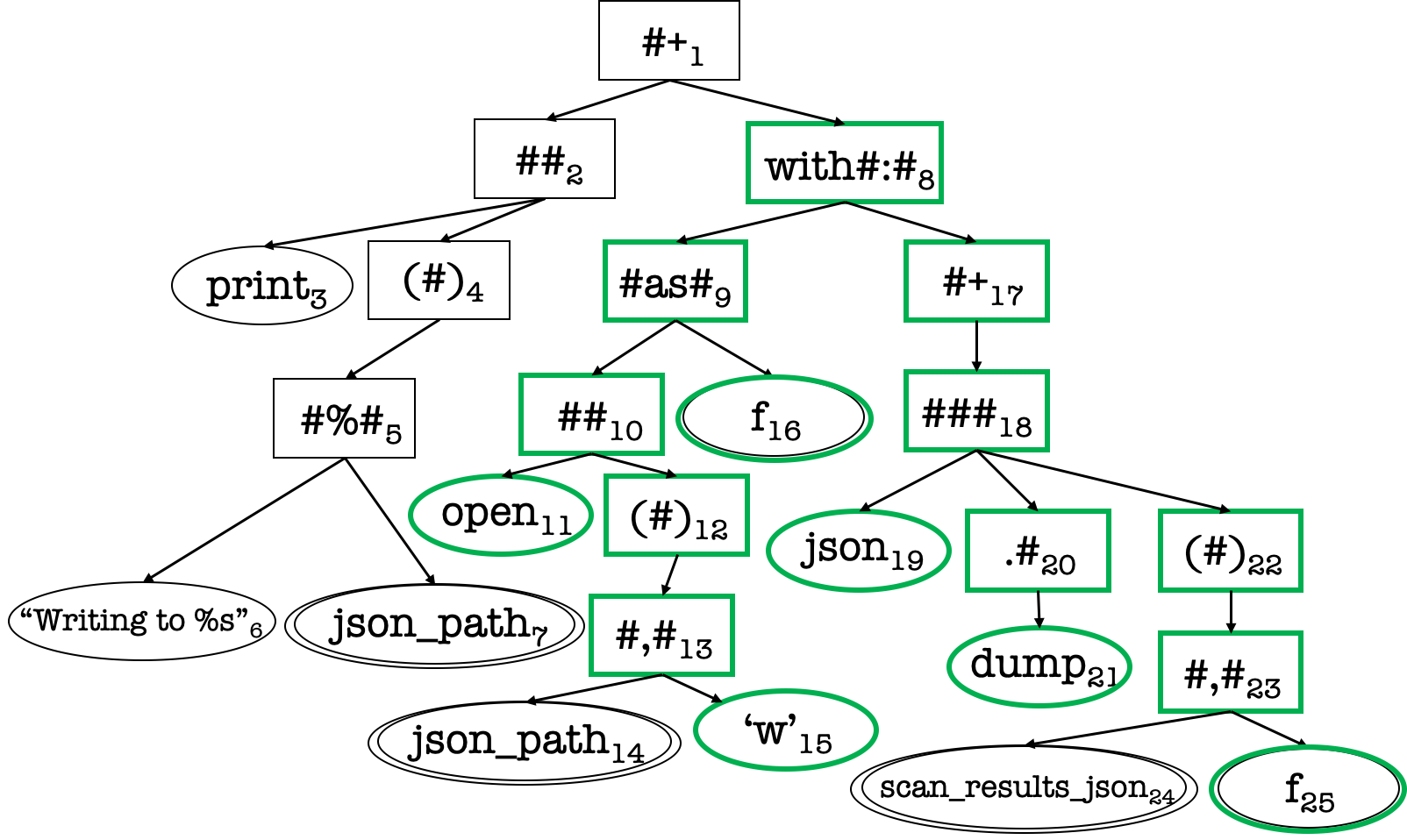}
\caption{The maximal pattern computed for the query \texttt{json.dump} from two different simiplified parse trees. The nodes in the pattern are highlighted in green.  The filler code is highlighted in blue.}
\label{fig:pattern_tree}
\end{figure}

Figure~\ref{fig:pattern_tree} shows the maximal pattern computed for the query \texttt{json.dump} from two simplified parse trees.  The nodes in the pattern are highlighted in green.

\begin{figure}[!ht]
\begin{algorithmic}
    \STATE{\textbf{Phase 1:}}
    \bindent
    \STATE{\textbf{Input:} a set of simplified parse trees $T$}
    \STATE{\textbf{Input:} the query tree $q$}
    \STATE{$T_c \leftarrow$ \{$t \in T \mid t$ contains $q$ as a subtree\}}
    \STATE{$t_c \leftarrow q$}
    \WHILE{no. of nodes in $t_c \leq \gamma$ and $|T_c| > |T| * \alpha$}
    \IF{$\exists (l, i, n_1, b)$ and $\exists n_2$ in nodes of $t_c$ such that \textsc{Support}(\textsc{Extend}($t_c, l, i, n_1, n_2, b$), $T_c$) $\geq$  \textsc{Support}(\textsc{Extend}($t_c, l', i', n_1', n_2', b$), $T_c$) for all $(l', i', n_1')$ and $n_2'$ in nodes of $t_c$}
    \STATE{$t_c \leftarrow$ \textsc{Extend}($t_c, l, i, n_1, n_2, b$)}
    \STATE{$T_c \leftarrow$ \{$t \in T_c \mid t $ contains $t_c$\}}
    \ENDIF
    \ENDWHILE
    \STATE{\textbf{return} $t_c, T_c$}
    \eindent
\end{algorithmic}
\caption{Phase 1 algorithm.}
\label{fig:phase1}
\end{figure}

\begin{figure}[!ht]
\begin{algorithmic}
    \STATE{\textbf{\textsc{Extend}$(t_c, l, i, n_1, n_2, b)$}}
    \bindent
    \STATE{Let $t_c = (N, L, E)$ }
    \STATE{$L \leftarrow L \cup \{n_1 \mapsto l\}$}
    \STATE{$N \leftarrow N \cup \{n_1\}$}
    \IF{$b$}
    \STATE{$E \leftarrow E \cup \{(n_2, i) \mapsto n_1\}$}
    \ELSE
    \STATE{$E \leftarrow E \cup \{(n_1, i) \mapsto n_2\}$}    
    \ENDIF
    \STATE{\textbf{return} $t_c$}
    \eindent
\end{algorithmic}
\caption{\textsc{Extend}$(t_c, l, i, n_1, n_2, b)$ adds the node $n_1$ with label $l$ to the node $n_1$ in $t_c$ and adds the edge $E(n_2,i) = n_1$ if $b$, and $E(n_1, i) = n_2$ otherwise.}
\label{fig:extend}
\end{figure}

\begin{figure}[!ht]
\begin{algorithmic}
    \STATE{\textbf{\textsc{Support}$(t, T)$}}
    \bindent
    \STATE{\textbf{return} $|\{t' \mid t' \in T \mbox{ and } t' \mbox{ contains t as a pattern}\}|$}
    \eindent
\end{algorithmic}
\caption{\textsc{Support}$(t, T)$ computes the number of trees in $T$ that contain $t$ as a pattern.}
\label{fig:support}
\end{figure}

\paragraph{Phase 2.} Once \aroma{} has computed a pattern contained in several trees, it tries to complete the pattern by adding the missing subtrees in the pattern.  In \aroma{} our goal is to show a real code snippet instead of a synthetic one, because we have found that programmers feel more confident with real code snippets. This means that we need to pick a minimal subtree from the final set $T_c$ such that the subtree contains the pattern.  We focus on a few properties of the tree which makes the common usage example code snippet short yet common.  First, the code snippet should be short.  Second, it should have more commonality to the code snippets in $T_c$.

In order to find the best subtree that extends the pattern found in Phase 1, we assign a \textbf{score} to each subtree obtained from the trees in $T_c$.  Let us call a subtree that can fill a missing subtree in the pattern a \textbf{filler tree}.
Let us also call the location at which a filler subtree is missing in the pattern a \textbf{hole}.  Therefore, a pattern has a fixed finite number of holes.  For a given tree $t$ in $T_c$ and a hole in the pattern, let $f$ be the filler that fills the hole in $t$. The score of $f$ is then the number of trees in $T_c$ where $f$ is the filler of the hole.  If the filler has more than $\beta_t$ of tokens (usually set to 5) or more than $\beta_c$ characters (usually set to 50), we set the score of $f$ to 0.  This ensures that the code snippets are concise.  We take the sum of all the fillers of the pattern in $t$ to compute the score of $t$.  We then pick the tree in $T_c$ which has the highest score and show its minimal subtree containing the pattern as the common usage example.
\aroma{} reconstructs the example with an in-order traversal of the subtree.
Figure~\ref{fig:pattern_tree} shows the filler code computed for the query \texttt{json.dump}.  The nodes in the filler code are highlighted in blue.



\subsection{Creating Multiple Common Usage Examples}

The algorithm above describes the process of generating a single common usage example for an API method.  However, in many scenarios a user may be interested in multiple yet diverse usage examples.  \aroma{} generates distinct usage examples for a single query as follows. \aroma{} generates the first common usage example using the regular \aroma{} algorithm described above---however, \aroma{} maintains a set of all the nodes added to the pattern in Phase 1. Let us call this set \textit{used\_nodes}.  \aroma{} then saves that pattern, and begins the example generation again with the same initial set of trees. However, if at any point the \emph{second} most common adjacent node is not in \textit{used\_nodes} and has at least half as many occurrences as the most common adjacent node, we add that node to the pattern instead. We then finish the example generation as normal. 

\aroma{} repeats this process $n$ times to create $n$ distinct usage examples. In the \aroma{} interface, we display the top three usage examples. 

\section{Evaluation}
\label{sec:evaluation}

We designed the following experiments to evaluate \aroma{}. In each experiment, we compared a code example of an API method generated by \aroma{} against a randomly selected code snippet containing the method from the code corpus. The random example displays the line of code where the method is called, as well as the two lines of code preceding and following the method call. This random example serves as a reasonable stand-in for an arbitrary code search result, as code search engines typically display 2-4 lines of additional context by default. We use this as a comparison point since code search is the de facto way developers learn APIs in real-world programming workflows---especially for proprietary APIs where no hand-written examples exist~\cite{brandt2010example, sadowski2015developers}. 

We aim to answer the following research questions:
\begin{questions}
  \item Do developers prefer \aroma{} code examples to code search results?
  \item How does \aroma{} perform against comparable tools on several quantitative metrics measuring code example quality?
  \item If \aroma{} is made accessible, will developers incorporate \aroma{} code examples into their workflows?
\end{questions}

\subsection{RQ1: Survey with Facebook Developers}

We first conducted a survey to measure the quality of code examples generated by \aroma{}, compared with code search results. The survey first displayed six common Python libraries, and asked participants to select two libraries they were most familiar with. The survey then showed ten API methods in each of the two selected libraries. For each method, two code examples were listed: the top-ranked example generated by \aroma{} (Option A) and a random example from the code search result (Option B). 
Participants were asked three questions about these two kinds of examples. Table~\ref{table:survey} shows the questions in the survey. 

We sent out the survey to 21 Facebook developers. 18 developers completed the survey (86\% response rate). 
Overall, the examples generated by \aroma{} were preferred over the random examples 97\% of the time. In addition, 66\% of participants agreed that it is helpful to see the number of code examples that follow the same API usage pattern. 100\% of participants agreed that it is helpful to color-code and distinguish code parts that are commonly shared among many examples.
When asked to describe what they liked and disliked about the two kinds of examples, participants expressed a markedly positive sentiment towards \aroma{}: one said, ``{\em the usage count is super useful especially to make sure that the code you are looking at is consistent with the rest of the codebase.}'' Another participant said, ``{\em I think that the formatting (color) makes it easier to quickly compare a few examples...and find the most relevant example for your use case.}''

\begin{table}[h!]
\caption{Questions asked in the Facebook survey.}
\label{table:survey}
 \begin{tabular}{|p{7.5cm}|} 
 \hline
 1. Suppose you were learning to use this library. Which code examples would you prefer to see? Select one:   
 \begin{itemize}
     \item Strongly prefer A
     \item Prefer A
     \item Somewhat prefer A
     \item Somewhat prefer B
     \item Prefer B
     \item Strongly prefer B
 \end{itemize} 
 \\ \hline
 2. To what extent do you agree with this statement: It is helpful to see the count of methods that contain a common usage pattern (e.g. “Common usage pattern found in 120 out of 2000 methods”).
  \begin{itemize}
     \item Strongly agree
     \item Agree
     \item Somewhat agree
     \item Somewhat disagree
     \item Disagree
     \item Strongly disagree
 \end{itemize} 
 \\ \hline 
 3. To what extent do you agree with this statement: It is helpful for a code example to be formatted so I can see what is common and what is unique to a specific use case (e.g. common part in black, unique part in gray).
   \begin{itemize}
     \item Strongly agree
     \item Agree
     \item Somewhat agree
     \item Somewhat disagree
     \item Disagree
     \item Strongly disagree
 \end{itemize} \\
 \hline
\end{tabular}
\end{table}

\subsection{RQ2: Quantitative Evaluation with Metrics}
\label{metrics}
In addition to the qualitative survey with real developers, we conducted a quantitative analysis of the quality of examples generated by \aroma{}. We defined several metrics to measure example quality:

\begin{itemize}
    \item {\bf\em Succinctness}: How many lines of code are in the example?
    \item {\bf\em Relevancy}: How relevant is the surrounding code in the example w.r.t. understanding the usage of the queried API?
    \item {\bf\em Representativeness}: How frequently do other examples in the code corpus follow the same pattern in the example?
\end{itemize}

{\em Succinctness} is measured by counting the number of lines in an example. We did not count empty lines or code comments. {\em Relevancy} is measured as the ratio of relevant lines in an example to total lines. A relevant line is a line whose meaning and connection to the query method is clear without additional explanation or context. 
Figure~\ref{fig:relevant_lines} illustrates this metric by showing random code search results and \aroma{} examples for two methods, with relevant lines bolded and the query methods highlighted. In the code search example for \texttt{np.array}, the first line does not show how or why \texttt{reshape} is called, so this line is deemed irrelevant. Without additional context, we also do not know what \texttt{discretize.EntropyMDL} does, so lines 4 and 5 are not relevant. In the \aroma{} example for \texttt{np.array}, lines 1 and 2 show calls to \texttt{np.array}, while lines 3 and 4 show the returned values of \texttt{np.array} being passed to \texttt{fit\_transform} -- so all of these lines are relevant. In the code search example for \texttt{pd.concat}, it is not clear what \texttt{df1} on lines 4 and 5 is used for, and how or if it pertains to \texttt{pd.concat} -- so these two lines are irrelevant. In the \aroma{} example for \texttt{pd.concat}, lines 1 and 2 show the initialization of variables passed to \texttt{pd.concat} in line 3 -- making all 3 lines relevant to understanding how \texttt{pd.concat} is used. {\em Representativeness} is measured by the ranking score that \aroma{} assigns to an example. Recall that this score is the sum of the number of occurrences of each filler option in the example. In this way, this score reflects how representative this example is of a common use of the query method. To measure the representativeness of the comparison baseline (i.e., code examples randomly selected from the original search result), we first check whether the method containing this random example is one of the methods containing the \aroma{} common usage pattern. If it is, we take that method's ranking score as the representativeness score. Otherwise, we assign the random example a representativeness score of 0. Notice that relevancy is measured by manually assessing the code snippets, while succinctness and representativeness are computed automatically. 

\begin{figure}[!ht]
\centering
\textbf{Code search examples}\par\medskip
\begin{Verbatim}[commandchars=\\\{\},frame=single]
	).reshape((100, 1))
\textbf{Y = \hl{np.array}([0] * 25 + [1] * 75)}
\textbf{table = data.Table.from_numpy(None, X, Y)}
disc = discretize.EntropyMDL()
dvar = disc(table, table.domain[0])
\end{Verbatim}
\begin{Verbatim}[commandchars=\\\{\},frame=single]
\textbf{Ozone1 = \hl{pd.concat}([df.Ozone] * K)}
\textbf{print(Time1.shape, Ozone1.shape,}
     \textbf{   Time1.describe(), Ozone1.describe())}
df1 = pd.DataFrame();
df1['Time'] = Time1.values;
\end{Verbatim}
\textbf{\aroma{} examples}\par\medskip
\begin{Verbatim}[commandchars=\\\{\},frame=single]
\textbf{X = \hl{np.array}(['a', 'b', 'c'])}
\textbf{y = \hl{np.array}([1, 0, 1])}
\textbf{out = encoders.JamesSteinEncoder(model='binary')}
                \textbf{.fit_transform(X, y)}
\end{Verbatim}
\begin{Verbatim}[commandchars=\\\{\},frame=single]
\textbf{sparse1 = pd.SparseSeries(val1, name='x')}
\textbf{sparse2 = pd.SparseSeries(val2, name='y')}
\textbf{res = \hl{pd.concat}([sparse1, sparse2], axis=1)}
\end{Verbatim}
\caption{Random code search examples and \aroma{} examples for several methods, with relevant lines bolded.\protect\endnote{These code snippets have been adapted from \url{https://github.com/renn0xtek9/Arithmos/blob/799fe071ab3a85ea9a0f86b8099548f11be96841/Arithmos/tests/test_discretize.py\#L111}, and \url{https://github.com/antoinecarme/pyaf/blob/master/tests/perf/test_ozone_long_series.py\#L23}. Accessed in March 2020.}}
\label{fig:relevant_lines}
\end{figure}

For this experiment, we considered four popular Python libraries: \texttt{Pandas}, \texttt{os}, \texttt{Numpy}, and \texttt{TensorFlow}. For each library, we selected the ten most used methods in GitHub -- forty methods total. We compared average succinctness, relevancy, and representativeness of the top \aroma{} example, a random code search result, and the top example from ProgramCreek~\cite{creek}. ProgramCreek is a website where users can query a Python library method and see functions from open source GitHub projects that call that method. Users vote on which functions represent the best example of a method. The top ProgramCreek example was taken to be the method with the most upvotes. Since the example was a complete method, relevancy was not a meaningful measurement; however, we were still able to measure length. 

Table~\ref{table:metrics-results} shows the quality of code examples generated by \aroma{}, randomly selected from code search results, and selected from ProgramCreek~\cite{creek}. Compared with examples from \aroma{} and ProgramCreek, random examples contained many more irrelevant lines of code, as well as long, uninformative identifier names. Meanwhile, ProgramCreek examples were on average over five times longer than \aroma{} examples. 

\begin{table}[ht]
\caption{The Quality of Code Examples for 40 Popular Methods in Python}
 \begin{tabular}{||c c c c||} 
 \hline
 Type of Example & Length & Relevancy & Representativeness \\ [0.5ex] 
 \hline\hline
 \aroma{} & 2.675 & .996 & 59.6 \\ 
 \hline
 Code Search & 3.9 & .640 & .2 \\
 \hline
 ProgramCreek & 13.8 & -- & -- \\
 \hline
\end{tabular}
\label{table:metrics-results}
\end{table}

We also collected 100 Hack API methods that had been queried in Facebook's code search website most frequently over a 30 day period. Hack is a programming language created by Facebook as a dialect of PHP~\cite{hack}. These 100 Hack methods were queried an average of 8.6 times, ranging from 5 times to 26 times. For these 100 methods, we measured the average length and representativeness of \aroma{} examples and random code search results. Since these 100 methods are proprietary API methods in Facebook, we were not able to find curated examples from ProgramCreek. As a result, we are not able to compare \aroma{} with ProgramCreek. 
Table \ref{table:metrics-results2} shows the quality of code examples generated by \aroma{} and randomly selected from code search results. Similar to the results on open-source libraries, examples generated by \aroma{} were significantly more concise and representative than examples selected from the original code search results. 

\begin{table}[ht]
\caption{The Quality of Code Examples for 100 Internal Methods in Facebook}
 \begin{tabular}{||c c c||} 
 \hline
 Type of Example & Length & Representativeness \\ [0.5ex] 
 \hline\hline
 \aroma{} & 3.5 & 116.6 \\ 
 \hline
 Code Search & 4.6 & 2.1 \\
 \hline
\end{tabular}
\label{table:metrics-results2}
\end{table}

\subsection{RQ3: Live Usage in Facebook}

We have integrated \aroma{} into Facebook's internal code search website. When users query a method name in Hack or Python, the top \aroma{} example is displayed first, before the standard code search results. There is a link to the full contents of the file containing the code snippet used in the example, and a "Show More Examples" button that displays two additional common usage patterns. 
\aroma{} only shows the three most common usage patterns, ensuring that its interface is compact and easy to use.
\aroma's integration into the code search platform was {\em frictionless}: developers began to use \aroma{} with no prior announcement or tutorial. 

\aroma{} is deployed on a dedicated set of servers to respond to queries from developers. Our search server has 24
cores, and on average takes 1.0 seconds end-to-end to generate Python code examples for the queries used in Section \ref{metrics}. The median response time is .8 seconds and the maximum is 2.3 seconds. \aroma{} re-indexes the millions of methods in Facebook's codebase daily. This indexing process works the same as in Aroma ~\cite{luan2019aroma}. On a 24-core server, this process takes 20 minutes on average. If \aroma{} were to be deployed on a larger codebase,  it would be possible to implement incremental indexing for only changed files. Since the goal of EG is to provide relevant and up-to-date usage examples, we show examples for only the most recently indexed version of the codebase, and we do not maintain past examples generated from prior versions. 

We have been logging the usage of \aroma{} in the code search website. We log each time a user copies or selects code from an \aroma{} example, clicks the file link, or clicks the "Show More Examples" button. Note that copying and selecting are the only events we log with a clear signal that the user actually reused code in the example. 

Over a period of 24 days, from April 20 to May 13, \aroma{} was triggered to generate code examples for an average of 1,171 code search queries per day. Facebook developers interacted with \aroma{} examples an average of 59 times per day, and copied or selected code from an \aroma{} example an average of 30 times per day. 
While this appears to be a low ratio of interactions to total examples generated, there are several factors to keep in mind. First, developers do not always query method names because they want to see code examples---for instance, a developer may instead be looking for a specific file or class. We have no way to determine what a developer's intentions are when they query a method name. Second, note that we integrated \aroma{} into the code search website without any public announcement. Therefore, Facebook developers may not even notice it among the other features of the code search website. The discoverability of \aroma{} is an orthogonal problem from its effectiveness, which we will investigate in the future.  
Finally, because a central feature of \aroma{} examples is succinctness, developers may be learning or "mentally copying" from an example without physically interacting with it. 

Despite these concerns, these results show that real developers indeed utilize \aroma{} in their workflows. A formal A/B test comparing \aroma{} examples against code search results remains as future work.




\subsection{Discussion}
\label{sec:discussion}
Evaluating example generation is an interesting and complicated problem. We initially attempted to design a study wherein developers receive a comprehensive list of API usage questions. They are asked to answer these questions for one API using \aroma{}, and for another using a realistic baseline of code search. The problem with this approach is that 
\aroma's focus is on providing a short list of idiomatic usage examples. \aroma{} makes no claim to offer the best or most informative usage example---but a succinct example representing a common usage pattern. Thus, we needed an evaluation that measured the benefit of seeing such a common usage pattern. 

We next attempted to design a human study wherein participants complete short programming tasks using \aroma{}. A main challenge was devising a control to measure \aroma{} against. An obvious candidate is Facebook's internal code search website. However, \aroma{} is not intended to replace code search altogether, but rather to be a complementary extension integrated into existing code search tools. Thus, it did not make sense to restrict the use of code search. However, Facebook employees are conditioned to use code search results as a go-to method for API inquiries, so even when the \aroma{} example contained salient, time-saving information, they often still wanted to page through code search results. Untangling what the user gained from \aroma{} versus what they gained from code search, or from documentation, proved difficult. In addition, success in solving a short programming task is extremely dependent on what background knowledge a developer has.

A tool like \aroma{} also runs the risk of identifying and perpetuating common anti-patterns. 
By indexing Facebook's codebase, we ensure that all code has been reviewed by a developer -- however, code can still become out of date or deprecated.
A potential solution would be to only index code in a codebase written after a certain threshold date. Another would be to indicate to the user in the UI the date when the code displayed in an example was written.
Exploring this issue further is left for future work.

\section{Related Work}
\label{sec:related}
Developers often search for code in their own codebases or online to fulfill programming needs such as learning new APIs and locating code snippets with desired functionality~\cite{sadowski2015developers,sim2011well, brandt2009two,umarji2008archetypal,montandon2013documenting}. For example, Sim et al.~conducted a lab study with 36 graduate students to evaluate the effectiveness of different code retrieval techniques~\cite{sim2011well}. In the demographic survey, 50\% of participants reported to search code online frequently and 39\% reported to search occasionally. Sadowski et al.~analyzed the search logs generated by 27 Google developers over two weeks~\cite{sadowski2015developers}. They found that developers issued an average of 12 code search queries per weekday.

There is a large body of literature in code search~\cite{Holmes2005, Mandelin2005, stylos2006mica, sahavechaphan2006xsnippet, thummalapenta2007parseweb, lemos2007codegenie, kim2010towards, brandt2010example, lazzarini2009applying, reiss2009semantics, Wang2010:matchingquery, mcmillan2011portfolio, mcmillan2012exemplar, kim2018f, gu2018deep, sirres2018augmenting, sivaraman2019active, yan2020code}. These techniques focus on 1) enriching search queries and 2) improving search algorithms. For example, beyond simple keyword descriptions, S$^6$~\cite{reiss2009semantics} and CodeGenie~\cite{lemos2007codegenie} allow users to identify relevant code based on test cases. Prospector~\cite{Mandelin2005} supports expressing type constraints such as desired input and output types in a query. Code-to-code search tools such as FaCoY~\cite{kim2018f} take code fragments directly as input and identify other similar code. Wang et al.~represented source code as a dependency graph to capture control-flow and data-flow dependencies in a program, and matched search queries against program dependence graphs~\cite{Wang2010:matchingquery}. Gu et al.~trained a neural network to predict relevant code examples given natural language queries~\cite{gu2018deep}.

Unlike our work, the aforementioned techniques provide limited support for browsing and assessing code search results. Previous studies have shown that it is cognitively demanding to navigate through code search results~\cite{duala2012asking, starke2009working}. As a result, 
developers often rapidly skim through a handful of search results and make a quick judgement about the quality of these results~\cite{brandt2009two}.  When browsing search results, they also often backtrack due to irrelevant or uninteresting information in search results~\cite{duala2012asking}. More specifically, Starke et al.~show that developers rarely look beyond five examples when searching for code examples~\cite{starke2009working}. These observations indicate that the code exploration process is often limited to a few search results, leaving a large portion of foraged information unexplored. 

Several approaches have been proposed to help developers navigate through code search results. To enable users to explore a large number of code examples simultaneously, Examplore constructs a code skeleton with statistical distributions of individual API usage features in those examples~\cite{glassman2018visualizing}. ALICE allows users to mark several search results as desired or undesired and then automatically filter the remaining search results, so users do not have to manually go through all of them~\cite{sivaraman2019active}. eXoaDocs employs program slicing to remove extraneous statements in a code example and then clusters sliced code examples based on the similarity of semantic characteristics such as invoked API methods in an example~\cite{kim2010towards}. Buse and Weimer improved eXoaDocs by synthesizing a single concise code example to summarize similar examples in a cluster~\cite{buse2012synthesizing}.  
 
Our approach differs from these techniques in several perspectives. While the tools described above rely on the syntax and semantics of the Java language, \aroma{} is language agnostic, requiring only a parser for the target language. 
Examplore requires a pre-defined API usage skeleton to register and align code examples, while \aroma{} does not require a pre-defined skeleton.
Buse and Weimer's tool generates usage examples for a target class, while \aroma{} generates usage examples for API or library methods.
Finally, to our knowledge, \aroma{} is the only tool designed to generate common usage patterns of APIs that has been integrated into the code search platform of a large software company, and is used by developers daily. 
 

\section{Conclusion}
\label{sec:conclusion}
We presented \aroma{}, a new tool for generating usage example for API methods. \aroma{} works by first indexing a large code corpus. Given a query method, it assembles a list of method bodies in the corpus containing that method, then finds the maximal subtree that contains the query API and is part of a meaningful proportion of methods. \aroma{} then reconstructs this subtree into a succinct, relevant and representative code example. 

To evaluate \aroma{}, we indexed a code corpus of 1.9 million Python methods, and designed a survey where we showed developers pairs of \aroma{} examples and code search results for commonly used methods in popular Python libraries. We observed that developers preferred \aroma{} examples to code search results 97\% of the time, and that 100\% of developers agreed that the color-coding of the common usage pattern in \aroma{} examples is helpful. Further, we defined several metrics to measure example quality, and quantitatively compared \aroma{} examples against code search results and ProgramCreek examples using these metrics. We found that across all metrics, \aroma{} performs better than these alternatives. Finally, we integrated \aroma{} into Facebook's internal code search website.  A log of developers' activities shows that developers indeed interact with \aroma{} examples. 

\theendnotes

\bibliographystyle{ACM-Reference-Format}
\bibliography{sample-base}

\end{document}